\documentclass[12pt]{iopart}

\usepackage{cite}
\usepackage{graphicx}
\usepackage{hyperref}
\usepackage{iopams}

\newcommand{\Wcm}{\,{\rm W \, cm^{-2}}}
\newcommand{\beq}{\begin{equation}}
\newcommand{\eeq}{\end{equation}}
\newcommand{\beqa}{\begin{eqnarray}}
\newcommand{\eeqa}{\end{eqnarray}}
\newcommand{\bma}{\mathbf}
\newcommand{\mum}{\,{\mu \rm m}}

\begin{document}

\title[Radiation Reaction Effects on Radiation Pressure Acceleration]{Radiation Reaction Effects on Radiation Pressure Acceleration}

\author{M. Tamburini$^{1}$, F. Pegoraro$^{1}$, A. Di Piazza$^{2}$, C. H. Keitel$^{2}$, and A. Macchi$^{3,1}$}

\address{$^{1}$Dipartimento di Fisica ``Enrico Fermi'', Universit\`a di Pisa, Largo Bruno Pontecorvo 3, I-56127 Pisa, Italy}\ead{matteo.tamburini@df.unipi.it}
\address{$^{2}$Max-Planck-Institut f\"ur Kernphysik, Saupfercheckweg 1, D-69117 Heidelberg, Germany}
\address{$^{3}$Istituto Nazionale di Ottica, CNR, research unit ``Adriano Gozzini'', Pisa, Italy}\ead{andrea.macchi@ino.it}

\begin{abstract}
Radiation reaction (RR) effects on the acceleration 
of a thin plasma foil by a superintense laser pulse in the radiation pressure dominated regime
are investigated theoretically.
A simple suitable approximation of the Landau-Lifshitz equation 
for the RR force and a novel leapfrog pusher for its inclusion 
in particle-in-cell simulations are provided.
Simulations for both linear and circular polarization of the laser pulse are performed and compared.
It is found that at intensities exceeding 
$10^{23} \Wcm$ the radiation reaction force strongly affects the dynamics 
for a linearly polarized laser pulse, 
reducing the maximum ion energy but also the width of the spectrum. 
In contrast, no significant effect is found for circularly polarized 
laser pulses whenever the laser pulse does not break through the foil. 
\end{abstract}

\submitto{\NJP}

\maketitle

\section{Introduction}

Present-day laser systems may deliver intensities up to $10^{22} \Wcm$ 
\cite{yanovskyOE08} at their focal spot. Even higher intensities 
of the order of 
$10^{24}-10^{26} \Wcm$ are envisaged at the Extreme Light Infrastructure (ELI).
Theoretical studies \cite{esirkepovPRL04} suggested
that in the interaction of a laser pulse with a thin foil,
Radiation Pressure Acceleration (RPA) becomes the dominant mechanism
of ion acceleration at intensities exceeding $10^{23} \Wcm$.
The radiation pressure dominated regime is attractive because of the
foreseen high efficiency and because of the quasi-monoenergetic features
expected in the ion energy spectrum.
Moreover, recent simulations suggest that multi-dimensional effects may allow a further
increase of the ion energy \cite{bulanovPRL10}.

At these extreme optical laser intensities $I \gtrsim 10^{23} \Wcm$, 
electrons become ultra-relativistic
within a fraction of the wave period experiencing super-strong accelerations and
therefore emitting relatively large amounts of electromagnetic radiation. Radiation
reaction (RR) is the influence of the electromagnetic field emitted by
each electron on the motion of the electron itself \cite{landau-lifshitz-RR} and
may become essential under the extreme conditions mentioned above. 
Early particle-in-cell (PIC) simulations \cite{zhidkovPRL02}
showed that RR effects become important
at intensities exceeding $5 \times 10^{22} \Wcm$ and increase nonlinearly
with the laser intensity.

In order to take RR effects self-consistently into account one should, 
in principle, solve the so-called Lorentz-Abraham-Dirac (LAD) equation
\cite{landau-lifshitz-RR}. It is well known that this equation is plagued 
by inconsistencies such as, for example, the appearance of ``runaway'' 
solutions 
in which an electron acquires an exponentially diverging acceleration even
without any external field. However, it has been shown that in the realm of
classical electrodynamics, i.e. neglecting quantum effects, the LAD equation
can be consistently approximated by the so-called Landau-Lifshitz (LL) equation
which is free from the mentioned inconsistencies 
\cite{landau-lifshitz-RR,spohnEPL00}.

In this paper, we investigate RR effects in the interaction
of a super-intense laser pulse with a thin foil in the RPA-dominant
or ``laser-piston'' \cite{esirkepovPRL04} regime by one-dimensional (1D)
PIC simulations both for linear and circular polarization.
Our approach is based on the LL equation of motion.
We identify leading terms in the LL equation and discuss suitable 
approximations. On this basis we develop a straightforward numerical 
implementation of the RR force in a standard PIC code. 
PIC simulations with RR effects included have been previously performed 
for various laser-plasma interaction regimes by several groups, either using 
an approach similar to the LL equation \cite{zhidkovPRL02} or using a 
different RR modeling 
\cite{kostiukovPoP03,kiselevPRL04,chenXXX09,naumovaEPJD09,naumovaPRL09,schlegelPP09}.

In our simulations, we check the RR's ability to reduce the electron 
heating which is responsible of the broadening of both the electron and ion 
spectrum. Indeed, recent studies for thick targets in the hole boring regime 
\cite{naumovaPRL09,schlegelPP09}
and ultrathin plasma slabs \cite{chenXXX09} suggested that the RR force
cools the electrons and may improve 
the quality of the accelerated ion bunches.
We found that in the linear polarization (LP) case,
the peak in the energy spectrum has both a lower
energy and a lower width when RR is included. At the same time,
the fraction of low energy ions is reduced.
However, strong modulations appear in the ion
energy spectrum after the acceleration phase both with and without RR
and eventually the quasi-monoenergetic features are lost.
In the circular polarization (CP) case, RR does not affect
the ion energy spectrum significantly 
even at intensities of the order of $10^{24} \Wcm$.
The differences between LP and CP appear to be related to the longitudinal 
electron oscillations driven by the $\bma{J} \times \bma{B}$ force 
in the LP case.
These oscillations allow a deeper penetration of the laser
pulse into the foil enhancing the effect of the RR force on electrons.
In the CP case, significant RR effects are found only for laser and 
target parameters such that  the laser pulse breaks through the foil due to 
nonlinear transparency, similarly to what was found in previous studies \cite{chenXXX09}.

\section{The Radiation Reaction force}

In classical electrodynamics, the effect of RR on the motion of an electron can
be taken included by means of an additional force besides the Lorentz force. 
The additional RR force basically describes the loss of energy and 
momentum by an accelerated electron which radiates EM waves, so that the 
electron trajectory
changes with respect to that predicted by the Lorentz force alone.
In the LL approach \cite{landau-lifshitz-RR} the RR force is written in a 
manifestly covariant form as
\beq 
\label{fR1}
f^\mu =
\frac{2e^3}{3mc^2} \left( \partial_\alpha F^{\mu\nu} u_\nu u^\alpha \right)
+ \frac{2e^4}{3m^2c^4} \left( F^{\mu\nu} F_{\nu\alpha} u^\alpha +
(F^{\nu\beta} u_\beta F_{\nu\alpha} u^\alpha) u^\mu \right)
\eeq
where $m$ and $e$ are the electron mass and charge respectively,
$u^\mu=(\gamma, \gamma \bma{v}/c)$ is its four-velocity and $F^{\mu\nu}$ is the 
electromagnetic tensor relative to the total electromagnetic field
acting on the electron except the field generated by the electron itself.

The importance of RR effects on the electron motion depends on the strength
and geometry of the EM fields, as well as on the electron energy which
is generally a function of the amplitude and of the frequency of the field 
itself. One would thus need to know at least the scaling of the electron 
energy with the laser pulse parameters for a preliminary evaluation of RR 
effects as well as for a discussion on the limits of validity of the chosen 
theoretical approach and on suitable approximations to it.
In the following discussion, we mostly refer to the case of the electron motion
in a plane wave. For this problem, the LL equation has an exact analytical 
solution for arbitrary pulse shape and polarization of the plane wave 
\cite{dipiazzaLMP08}. Such solution thus 
provides a useful benchmark and reference for RR effects in superstrong laser 
fields. In a many-particle system such as a high-density plasma, the collective 
EM fields are generally much more complicated but the plane wave 
results may provide some guidance for their interpretation.

We first recall that the LL approach is classical and quantum electrodynamics 
effects are neglected. In the interaction between an intense laser field (with 
peak intensity $I$ and wavelength $\lambda$) and an ultra-relativistic electron 
(with Lorentz factor of the order of $\gamma$) 
this is in general allowed if $\gamma\sqrt{I/I_{cr}}\ll 1$ 
and $\gamma \lambda_c / \lambda \ll 1$ \cite{landau-lifshitz-RR}, 
where $I_{cr}=cE^2_{cr}/8\pi \approx 2.3\times 10^{29}\;\Wcm$ is the intensity 
corresponding to the critical field $E_{cr}=m^2c^3/\hbar |e|$ of quantum 
electrodynamics \cite{ritusJSLR85} and 
$\lambda_c = \hbar /mc \approx 3.9 \times 10^{-7} \mum$
is the Compton wavelength.
These conditions ensure that the momentum of the photons emitted or absorbed by 
the electron is negligible.
Moreover, the force related to the electron spin might not be negligible in
comparison to the RR force. 
In fact, the dynamics of a particle with a spin degree of freedom
in an external electromagnetic field
can be described in the classical framework by the Frenkel force
\cite{walser-keitelLMP01} (see also \cite{grallaPRD09} for a different 
derivation of both the RR and the spin force)
\beq  \label{fSpin}
f^\mu_S = - \frac{1}{2} Q^{\gamma\delta} \partial^\mu F_{\gamma\delta}
+ \frac{1}{2} \left(Q^{\gamma\delta} \partial_\lambda F_{\gamma\delta} u ^\lambda \right) u^\mu
\eeq
where $Q^{\gamma\delta} = \varepsilon^{\gamma\delta\alpha\beta} u_\alpha m_\beta$,
$m^\alpha$ is the magnetic dipole moment four-vector and 
$\varepsilon^{\gamma\delta\alpha\beta}$ is the Levi-Civita symbol 
($\varepsilon^{0123}=+1$).
The analysis of the case of a plane wave 
(electric field amplitude $E$, central frequency $\omega$
and pulse length $\tau$) shows that the spin force
is about $\sim \gamma/\alpha \simeq 137\,\gamma$ times the term in the LL force 
(\ref{fR1}) containing the derivatives 
of the field tensor, i.e. the term proportional to  
$\partial_\lambda F^{\mu\nu}$
(here $\alpha = e^2 / \hbar c \approx 1/137$ is the fine-structure constant). 
However, it can also be shown that
the spin effects remain smaller than those due to the last term in 
Eq.(\ref{fR1}) if $\alpha a_0\omega\tau \gtrsim 1$ where $a_0=|e|E/m\omega c$
(the effect of the last RR term cumulates with time). 
Since $\tau>2\pi/\omega$ and $a_0>300$ in our simulations the
latter condition is well satisfied.
It is therefore consistent, in a regime where RR effects are relevant and
quantum effects are subdominant, to neglect
\emph{both} the spin force and the first term of the RR force in Eq.(\ref{fR1}).

The PIC simulations with the RR force included are performed in the 
laboratory frame, i.e. the frame where the plasma target is initially at rest.
In the laboratory frame we write down the LL equation in three-dimensional,
non-manifestly covariant form as
\begin{eqnarray}
\frac{d \mathbf{p}}{d t} & = & - \Big(\mathbf{E} + \mathbf{v} \times \mathbf{B} \Big) \nonumber \\
& - & \left( \frac{4}{3}\pi \frac{r_e}{\lambda} \right) \gamma \Big[ \Big( \frac{\partial}{\partial t}
+ \mathbf{v} \cdot \nabla \Big) \mathbf{E} + \mathbf{v} \times \Big( \frac{\partial}{\partial t}
+ \mathbf{v} \cdot \nabla \Big) \mathbf{B} \Big] \nonumber \\
& + & \left( \frac{4}{3}\pi \frac{r_e}{\lambda} \right)
\Big[ \Big( \mathbf{E} + \mathbf{v} \times \mathbf{B} \Big) \times \mathbf{B} +
\Big( \mathbf{v} \cdot \mathbf{E} \Big) \mathbf{E} \Big]  \nonumber \\
& - & \left( \frac{4}{3}\pi \frac{r_e}{\lambda} \right) \gamma^2
\Big[ \Big( \mathbf{E} + \mathbf{v} \times \mathbf{B} \Big)^2
- \Big( \mathbf{v} \cdot \mathbf{E} \Big)^2  \Big] \mathbf{v}   \label{3dLL}
\end{eqnarray}
where $\mathbf{p}$ is the electron momentum, $r_e \equiv e^2 / m c^2 \approx 2.8 \times 10^{-9} \mum$
is the classical electron radius, $\lambda=2\pi c/\omega$ is the laser wavelength
and we use dimensionless quantities as in the PIC code.
Time is normalized in units of $\omega^{-1}$, 
space in units of $c \omega^{-1}$, momenta in units of $m c$.
Consequently, EM fields are normalized in units of $m \omega c/|e|$
and densities in units of the critical density $n_c=m\omega^2/4\pi e^2$.
The first RR term of Eq.(\ref{3dLL}), i.e. the one containing the 
``total'' time derivative of the EM fields, corresponds to the negligible
term in the manifestly covariant LL Eq.(\ref{fR1}) and is reported here for 
completeness but neglected in the calculations for the above explained reasons.

Since RR effects are important for ultra-relativistic electrons $\gamma \gg 1$, 
the last term in Eq.(\ref{3dLL}) (proportional to $\gamma^2$) dominates over 
the preceding one.
From a practical point of view, the smaller term may often be neglected even 
though
the on-shell condition $u_{\mu}u^{\mu}=1$ is lost neglecting this term.
Although single particle and PIC tests with and without this term showed no 
significant difference, both terms were included in our PIC simulations. 
It is possibly instructive however to neglect for a moment the smaller 
term and write down an effective reduced LL equation in the \emph{lab frame}

\beq \label{redLL}
\frac{d \mathbf{p}}{d t} = \mathbf{f}_L - d \, \mathbf{v}
\eeq
where $\mathbf{f}_L \equiv -(\mathbf{E} + \mathbf{v} \times \mathbf{B})$
and $d$ is given by
\beqa \label{damping}
d & \equiv & \left( \frac{4}{3}\pi \frac{r_e}{\lambda} \right) \gamma^2
\Big[ \mathbf{E}^2 - \left( \mathbf{v} \cdot \mathbf{E} \right)^2 + 
\mathbf{v}^2 \mathbf{B}^2 - \left( \mathbf{v} \cdot \mathbf{B} \right)^2
- 2 \mathbf{v} \cdot \left( \mathbf{E} \times \mathbf{B} \right) \Big]  \nonumber \\
& = & \left( \frac{4}{3}\pi \frac{r_e}{\lambda} \right) \gamma^2 \left[\mathbf{f}_L^2 - (\mathbf{v} \cdot \mathbf{f}_L)^2\right] \geq 0.
\eeqa
In Eq.\eref{damping}, RR effects appear as a ``friction'' term
with a nonlinear and anisotropic friction coefficient given by $d$.
When an electron ``crosses'' an EM field, it feels a viscous force opposite
to its velocity.

For an ultrarelativistic electron, the friction coefficient $d$ 
may be used as a measure of the strength of the RR force in units of 
$m \omega c$. In the case of motion in a plane wave, $d$ may be 
compared directly to the normalized wave amplitude $a_0$.
Setting $\mathbf{E} \times \mathbf{B}$ along the positive $x$ axis, 
the RR force vanishes ($d \rightarrow 0$) when $v_x \rightarrow 1$,
has its maximum value 
($d \rightarrow \left(\frac{4}{3}\pi\frac{r_e}{\lambda}\right)\gamma^24a_0^2$)
when $v_x \rightarrow -1$ and finally
$d\rightarrow\left(\frac{4}{3}\pi\frac{r_e}{\lambda}\right)\gamma^2a_0^2$ when 
$(v_y^2+v_z^2)\rightarrow 1$.

The friction effect of the RR physically corresponds to the incoherent 
emission of high frequency radiation by ultrarelativistic electrons.
When the RR is included in the numerical simulation of a collisionless, 
relativistic plasma, it is typically not feasible to resolve electromagnetic 
waves at such high frequencies, much larger than the inverse of the temporal 
resolution. Thus, it is assumed that such radiation escapes from the system 
without re-interacting with other electrons. Notice that even a solid-density 
plasma is transparent to such radiation, since in the present regime the RR 
effect is mostly due to the emission of radiation with photon energies in the 
MeV range, while the plasma frequency corresponds to at most a few hundreds of 
eV. From the point of view of energy balance, then, the energy radiated at high 
frequencies appears as a loss term or ``dissipation''. The percentage of 
radiative loss is measured by comparing the energy balance simulations 
including RR with simulations without RR, where the total energy of fields 
and particles is conserved within the limits of numerical accuracy (typically 
within $~1\%$ in our PIC code).

It may be worth recalling that, for what concerns the LL equation of 
motion, energy and momentum are not conserved exactly for the single electron. 
This is due to some terms which are neglected when deriving the LL equation 
from the LAD equation under the assumption that the radiation force in the 
instantaneous rest frame of the electron is much smaller than the Lorentz force
\cite{landau-lifshitz-RR}.
However, the neglected terms are much smaller than quantum corrections 
\cite{landau-lifshitz-RR}, thus the approximation is consistent with a classical
treatment. In a different approach to the RR force recently presented
in Ref.\cite{sokolovPoP09}, a different couple of classical equations of motion
are derived phenomenologically starting from the requirement of energy-momentum 
conservation of the system of the electromagnetic field plus the 
radiating electron.

\section{The numerical approach}

Our PIC code is based on the standard, widely used Boris particle pusher 
\cite{birdsall-langdon} and leap-frog schemes to advance and accelerate particles.
We developed a simple numerical scheme to insert the RR force in the PIC code while
keeping the standard particle pusher for the Lorentz force unchanged.
As will be clear below, this scheme is based on the assumption that the
acceleration of particles is dominated by the Lorentz force, with the RR 
force giving a smaller, albeit non negligible contribution.

We write the total force $\mathbf{f}$ acting on the electron
as the sum of two forces $\mathbf{f}_L$ (already introduced) and $\mathbf{f}_R$,
with
\begin{equation}
\label{f_r}
\mathbf{f}_R = -\left( \frac{4}{3} \pi \frac{r_e}{\lambda} \right)
\Big\{ \mathbf{f}_L \times \mathbf{B} - ( \mathbf{v} \cdot \mathbf{E} ) \mathbf{E} +
\gamma^2 \Big[ \mathbf{f}_L^2 - ( \mathbf{v} \cdot \mathbf{E} )^2  \Big] \mathbf{v}\Big\}.
\end{equation}
Then, the equation of motion of the electron reads
\beq
\frac{d\mathbf{p}}{dt} = \mathbf{f} = \mathbf{f}_L + \mathbf{f}_R.
\eeq
Assuming that forces and momenta are known at integer and half-integer
timesteps respectively, the full leap-frog step is
\beq \label{tot}
\frac{\mathbf{p}^{(n+1/2)} - \mathbf{p}^{(n-1/2)}}{\Delta t} = \mathbf{f}^{(n)} = \mathbf{f}^{(n)}_L + \mathbf{f}^{(n)}_R
\eeq
where $\Delta t$ is the timestep.
Now, we consider the leap-frog step for two ``helper'' momenta ${\bf p}_L$
and ${\bf p}_R$
\beq \label{part1}
\frac{\mathbf{p}^{(n+1/2)}_L - \mathbf{p}^{(n-1/2)}_L}{\Delta t} = \mathbf{f}^{(n)}_L, \quad
\frac{\mathbf{p}^{(n+1/2)}_R - \mathbf{p}^{(n-1/2)}_R}{\Delta t} = \mathbf{f}^{(n)}_R
\eeq
and assume
$\mathbf{p}^{(n-1/2)}_L = \mathbf{p}^{(n-1/2)}_R = \mathbf{p}^{(n-1/2)}$.
Thus, from the above equations we easily obtain
\beq \label{fullstep}
\mathbf{p}^{(n+1/2)} = \mathbf{p}^{(n+1/2)}_L + \mathbf{p}^{(n+1/2)}_R - \mathbf{p}^{(n-1/2)}
\eeq
This means that, starting at time $t^{(n)}$ and position $\mathbf{x}^{(n)}$ with 
$\mathbf{p}^{(n-1/2)}$, firstly $\mathbf{p}^{(n+1/2)}_L$ and  $\mathbf{p}^{(n+1/2)}_R$
are calculated \emph{independently} using $\mathbf{f}^{(n)}_L$ and $\mathbf{f}^{(n)}_R$
respectively, and finally Eq. (\ref{fullstep}) is employed 
to obtain the full leap-frog step $\mathbf{p}^{(n+1/2)}$.
It is worthwhile noticing that this is a {general} result as
we have used only the superposition property
of the force without any assumption about $\mathbf{f}_L$ and $\mathbf{f}_R$.

The previous algorithm allows to keep the standard leap-frog pusher for
the Lorentz force and to develop an independent pusher for the RR force alone.
Using Eq. (\ref{part1}) we can recast Eq. (\ref{fullstep}) as
\beq \label{pertfull}
\mathbf{p}^{(n+1/2)} = \mathbf{p}^{(n+1/2)}_L + \mathbf{f}^{(n)}_R \Delta t = \mathbf{p}^{(n+1/2)}_R + \mathbf{f}^{(n)}_L \Delta t
\eeq
Now, in order to compute the momentum change from step $n-1/2$ to $n+1/2$ 
due to the Lorentz and the RR force, an estimate of the electron's velocity at halfstep $n$ is needed.
To this aim, we first advance $\mathbf{p}^{(n-1/2)}$ to $\mathbf{p}^{(n+1/2)}_L$
using the Boris pusher for the Lorentz force, then we use $\mathbf{p}^{(n+1/2)}_L$ to estimate the
\emph{total} momentum $\mathbf{p}^{(n)}$ and velocity $\mathbf{v}^{(n)}$ at half time step as
\beq  \label{estim1}
\mathbf{p}^{(n)} \approx \frac{\mathbf{p}^{(n+1/2)}_L + \mathbf{p}^{(n-1/2)}}{2};
\qquad
\mathbf{v}^{(n)} \approx \frac{\mathbf{p}^{(n)}}{\gamma^{(n)}}
\eeq
where
\beq  \label{estimgamma}
\gamma^{(n)} = \sqrt{1 + \left( \mathbf{p}^{(n)} \right)^2}
\eeq
Next we use Eqs. (\ref{estim1}) and (\ref{estimgamma}) together with the fields
$\mathbf{E}^{(n)}, \; \mathbf{B}^{(n)}$ at half time-step to compute the full 
term
$\mathbf{f}^{(n)}_R$ according to Eq.~\eref{f_r}. This task is particularly 
simple because many terms of $\mathbf{f}_R$
can be written by $\mathbf{f}_L$ directly (see Eq.~\eref{f_r}).

This particle pusher was tested comparing the numerical results
for a single electron in a monochromatic 
plane wave both with the known analytical solution
\cite{dipiazzaLMP08} and with the numerical solution obtained using a
4th order Runge-Kutta scheme. 
These numerical calculations confirmed that the inclusion of the 
RR force according to the above method preserves the accuracy and stability
of the standard Boris pusher algorithm.
The range of intensities in the tests was from $10^{22} \Wcm$ to $10^{24} \Wcm$
with $\lambda = 0.8 \mum$.
Taking as an example case an electron with initial momentum 
$p_{x_{0}} = - 200 \, m c$ and a wave with $a_0 = 350$ and $\lambda = 0.8 \mum$,
we found our particle pusher to yield a phase error in the longitudinal 
momentum of $\sim 0.1(2\pi/\omega)$ after a run time of $500\,\omega^{-1}$ 
using a timestep $\Delta t=0.01\,\omega^{-1}$. 
The corresponding relative error in the 
displacement in the direction of wave propagation was $\sim 4\times 10^{-4}$.
The one-particle tests were performed using
the complete expression of the LL force (\ref{3dLL}) with the fields
and their derivatives as given functions of space and time.
These tests also confirmed that the derivative term in the LL
force \eref{3dLL} is negligible. 
The inclusion of the RR force in the PIC code according to the above 
described approach leads to approximately a 10\% increment of the computing 
time.

\section{The PIC simulations}

We performed PIC simulations with a plasma slab of ions (protons) with 
uniform initial density $n_0$. 
Since our primary aim is to evaluate the importance of RR effects on 
laser-plasma dynamics and ion acceleration in the regime of radiation pressure 
dominance, we restrict ourselves 
to a one-dimensional (1D) geometry for the sake of
simplicity and the possibility of using high numerical resolution. 
Multi-dimensional effects, which may be important to determine the features of 
ion acceleration in this regime \cite{bulanovPRL10}, will be presented in 
forthcoming publications; preliminary 2D simulations \cite{tamburiniNIMA11}
show qualitatively similar trends to the 1D case. It is worth
noticing that, as the momentum space in the 1D PIC code is already 
three-dimensional, our numerical approach can be readily implemented in 
a multi-D code employing the same particle pusher. The modest increase in 
computational time implied by our method might be essential to be able to 
perform large-scale multi-D simulations with RR included.

We report results for a laser intensity $I = 2.33 \times 10^{23}\Wcm$ and a
laser wavelength $\lambda = 0.8 \mum$, corresponding to a dimensionless 
parameter $a_0=328$. 
In all the simulations, the density $n_0 = 100 n_c$ 
and the profile of the laser field amplitude has a ``trapezoidal'' shape in 
time with one-cycle, $\sin^2$-function rise and fall and a five cycles 
constant plateau. 
The laser pulse front reaches the edge of the plasma foil at $t=0$.
The foil thickness is $\ell = 1 \lambda$ in all the simulations except 
for the ``transparency'' case reported below, for which $\ell = 0.3 \lambda$.
We considered both Circular (CP) and Linear (LP) polarization of the laser
pulse. 
The parameters are similar to those of the 3D simulations in
Ref.\cite{esirkepovPRL04} where the laser pulse was linearly polarized.
According to \cite{esirkepovPRL04}, RPA dominates 
the acceleration of ions in the plasma foil when the laser intensity 
$I \gtrsim 10^{23} \Wcm$.
To our knowledge, neither RR effects nor CP have been studied so far in 
such a regime of laser and plasma parameters. 
The effects of CP have been studied
extensively at lower intensities (see e.g. Ref.\cite{macchiNJP10} 
and references therein) showing that, with respect to LP, the use of CP 
quenches the generation of highly relativistic electrons making RPA dominant
also at such lower intensities.
Concerning RR effects, in Ref.\cite{esirkepovPRL04} it was suggested 
that the higher velocity the plasma foil is accelerated to, the lower the RR
force becomes because of the relativistic increase of the laser wavelength 
$\lambda'$ in the foil frame, making the RR strength parameter 
$\sim r_e/\lambda'$ increasingly small.
The expected quenching of RR effects may also be explained with the 
help of the ``reduced'' LL equations \eref{redLL}-\eref{damping}: when the foil
moves coherently with a velocity close to $c$, the amplitude of the reflected
wave is strongly reduced at any time in the laboratory frame; thus, 
the electrons at the surface of the foil can be considered as moving
with a velocity $v_x \simeq c$ in the field of the incident plane wave and 
parallel to its propagation direction, and the RR force almost vanishes.

\begin{figure}[th!]
\includegraphics[width=0.48\textwidth]{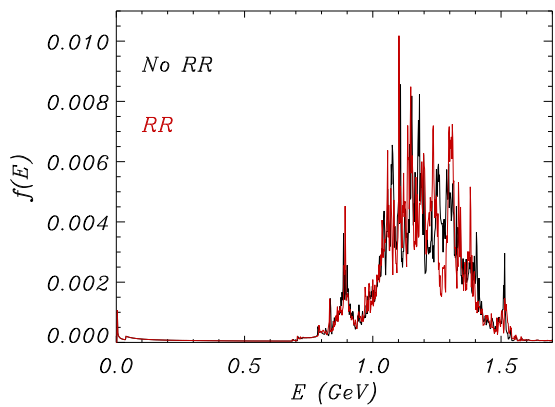}
\caption{\label{CPspectrum} Ion energy spectrum $f(E)$ 
at $t = 46 \, T$ with (red) and without (black) RR for CP. 
The laser intensity is $I = 2.33 \times 10^{23} \Wcm$ and the target thickness 
is $\ell=1\lambda$. See the text for the parameters common to all the 
simulations.}
\end{figure}

\begin{figure}[h!t]
\includegraphics[width=0.48\textwidth]{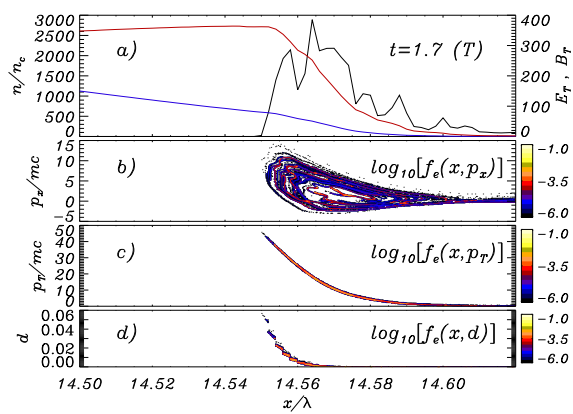}
\caption{\label{CPplot} Snapshot at $t = 1.7 \, (T)$
of the ``skin''layer of the foil for CP and $I = 2.33 \times 10^{23} \Wcm$.
The foil was initially placed between $x=14\, \lambda$ and $x=15\, \lambda$. 
a) The electron density (black),
the modulus of the transverse electric $E_\perp =\sqrt{E_y^2+E_z^2}$ (blue) 
and magnetic $B_\perp =\sqrt{B_y^2+B_z^2}$ (red) fields.
Distribution of b) longitudinal momentum $p_x$, c) modulus of the transverse momentum 
$p_\perp =\sqrt{p_y^2+p_z^2}$ and d) friction coefficient $d$.}
\end{figure}

In the CP case, we found that RR effects
on the ion spectrum (distribution of protons per unit energy)
are negligible as shown in Fig.~\ref{CPspectrum} for a time 
$t=46T$ where $T=\lambda/c$ is the laser period.
Even at higher intensities, RR effects on the ion 
spectrum are weak provided that there is not a strong 
transmission of the laser pulse through the foil.
In the simulation corresponding to Fig.~\ref{CPspectrum}, the laser pulse
penetrates into the plasma for a small distance of the order of $\lambda/20$,
and the fields in the plasma are much smaller than the fields in vacuum.
As a consequence, the friction coefficient $d$ introduced in Eq. (\ref{damping})
is very small compared to $a_0$. 
The spatial profiles of both the fields and the coefficient $d$ in the 
``skin''layer are shown in Fig.~\ref{CPplot} 
The order of magnitude of the normalized transverse momentum is
$\bma{p}_\perp \sim 10$ and of the friction coefficient is $d \sim 10^{-2}$.
It is worth mentioning that Fig.~\ref{CPplot} shows a snapshot at 
$t = 1.7 \, (T)$
but the typical values of the friction coefficient $d$ are always of the
same order of magnitude for CP. In contrast, for LP the friction 
coefficient $d$ attains much larger values at the same instant, as 
discussed below. 
We also notice that, for CP, we obtain qualitatively similar results
also at higher intensities, up to $10^{24} \Wcm$. However, at such extremely
high intensities the condition of validity of the classical approach
($\gamma\sqrt{E/E_{cr}}<1$) tends to be violated, so at least such results 
should be taken with caution and an analysis based on quantum RR effects
might be necessary.

\begin{figure}[th!]
\includegraphics[width=0.48\textwidth]{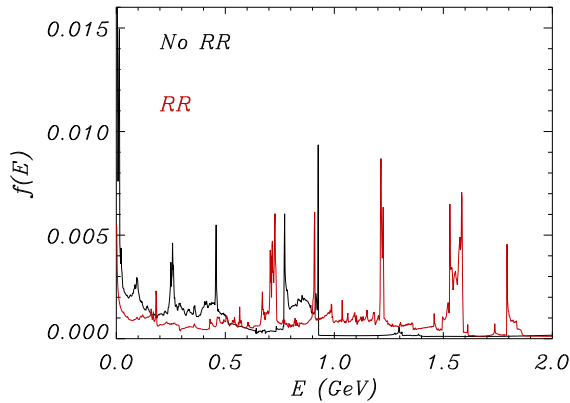}
\caption{\label{CPtransp}  Ion energy spectrum $f(E)$ at 
$t = 46 \, (T)$ for a simulation with the same
parameters as Fig.~\ref{CPspectrum} but with a target thickness 
$\ell=0.3\lambda$. In this case the laser pulse breaks through the foil
and RR effects are evident.}
\end{figure}

Reducing the foil density or thickness, the laser pulse
may break through the foil.
In this case more electrons move in a strong electromagnetic field
becoming ultrarelativistic in a fraction of wavecycle and RR effects 
strongly affect the ion spectrum, as shown in Fig.~\ref{CPtransp}.
In particular, when RR is included, peaks in the energy spectrum appear at 
energies higher than in the case without RR. This result is similar to that 
obtained in Ref.\cite{chenXXX09} at lower intensities ($\sim 10^{22} \Wcm$ ), 
where it was suggested that RR effects ``improve'' the ion spectrum in the
optical transparency regime. Our explanation is that the effective 
``dissipation'' due to RR leads to a later breakthrough of the laser pulse 
through the foil, favoring a longer and more efficient RPA stage. 
However, comparing  Fig.~\ref{CPtransp}
with the thicker target case in Fig.~\ref{CPspectrum}, it is evident that 
the spectrum becomes very far from monoenergetic, while the maximum ion
energy increases only slightly.
Hence in our simulations ``optimal'' conditions for ion acceleration are 
found for the case of Fig.~\ref{CPspectrum}; for the corresponding laser 
and plasma parameters, RR effects are negligible.

\begin{figure}[h!t]
\includegraphics[width=0.48\textwidth]{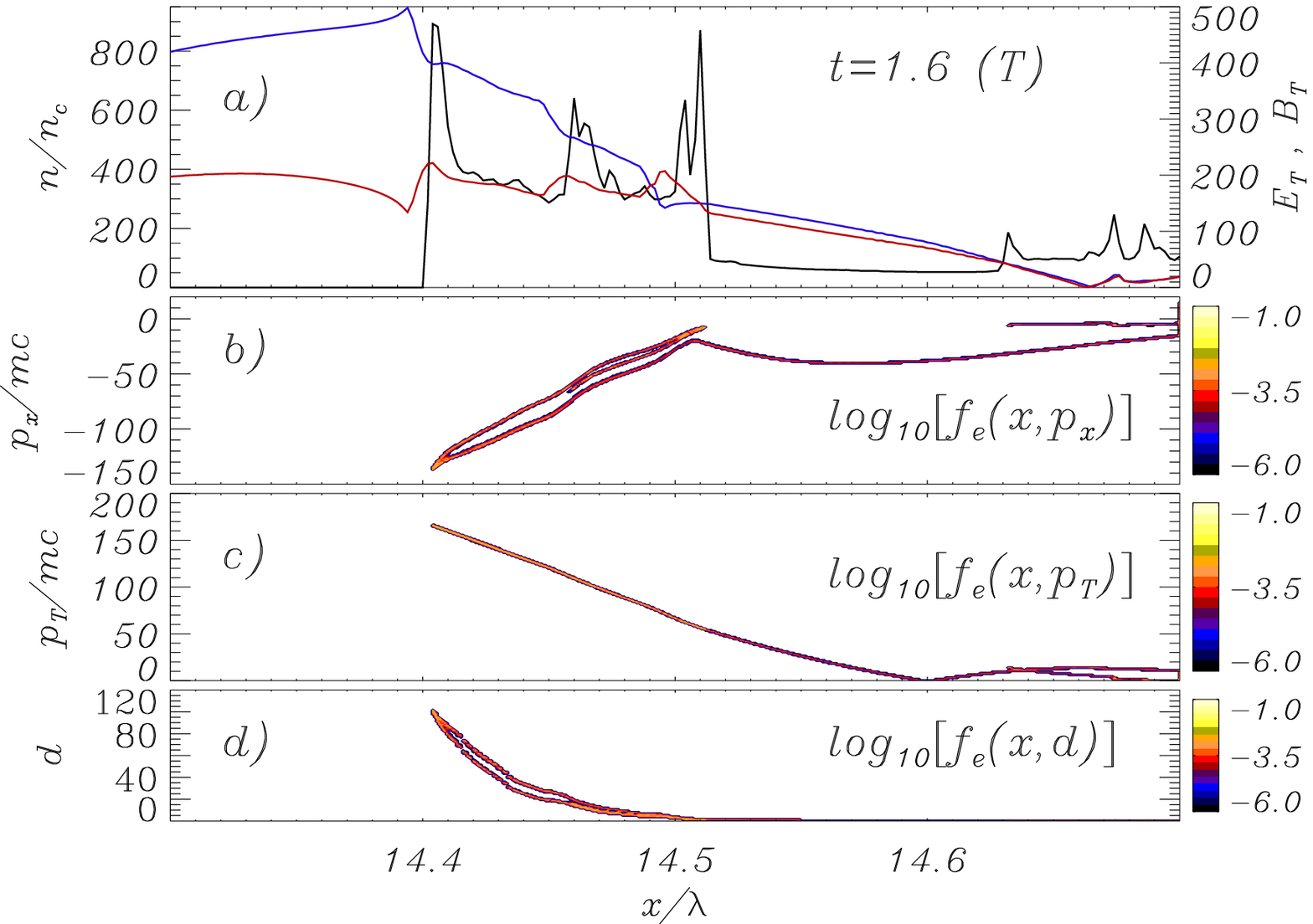}
\includegraphics[width=0.48\textwidth]{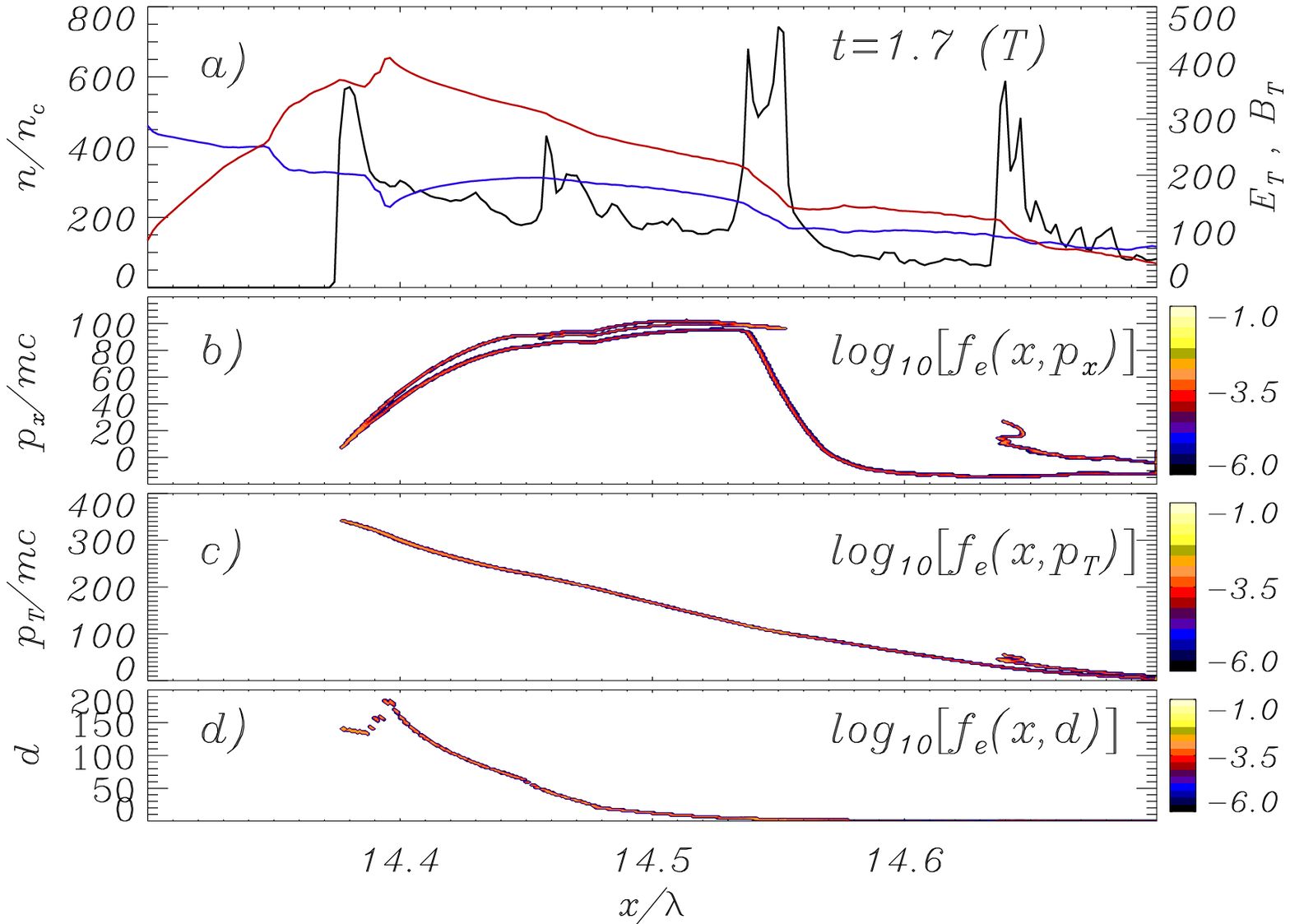}
\caption{\label{LPplot} Snapshot at $t = 1.6 \, (T)$ and $t = 1.7 \, (T)$
of the ``skin''layer of the foil for LP and $I = 2.33 \times 10^{23} \Wcm$.
The strong longitudinal oscillations driven by the $\mathbf{J}\times\mathbf{B}$ 
force allow a deeper penetration of the laser pulse into the foil compared to 
the CP case.
The foil was initially placed between $x=14\, \lambda$ and $x=15\, \lambda$. 
a) The electron density (black), the modulus of the transverse electric 
$|E_y|$ (blue) and magnetic $|B_z|$ 
(red) fields. Distribution of the b) longitudinal momentum $p_x$, c) 
modulus of the transverse momentum $p_\perp =|p_y|$
and d) friction coefficient $d$. Notice the change of the scale from
$t = 1.6 \, (T)$ and $t = 1.7 \, (T)$ in the frames b), c) and d).
We remark that the longitudinal momentum $p_x$ distribution changes
of orders of magnitude in $0.1 \, (T)$ due to the $\bma{J} \times \bma{B}$ 
force.}
\end{figure}

In the LP case the foil is accelerated by radiation pressure too but, unlike 
the CP case, the laser pulse does penetrate up to a fraction
of the order of $\lambda/4$ at the front surface of the foil, 
as shown in Fig.~\ref{LPplot}. 
The two snapshots are selected both to show values of $d$ close to its
maximum in time and to make a direct comparison with the CP case of
Fig.~\ref{CPplot}. It is found that a larger fraction of 
electrons at the front surface move in a strong electromagnetic field of the 
same order of the vacuum fields. In this case, the friction coefficient 
function $d$ reaches values of $d \approx 10^2$ (Fig.~\ref{LPplot}) which are 
comparable with the Lorentz force ($a_0 = 328$). 
The deeper penetration of the laser pulse is correlated with 
the strong longitudinal oscillatory motion driven by the 
oscillating component of the $\bma{J} \times \bma{B}$ force 
which is suppressed for CP. Large numbers of electrons are pushed
periodically inside the foil producing strong
fluctuations of the electron density (see Fig.~\ref{LPplot} part a)).

\begin{figure}[h!t]
\includegraphics[width=0.48\textwidth]{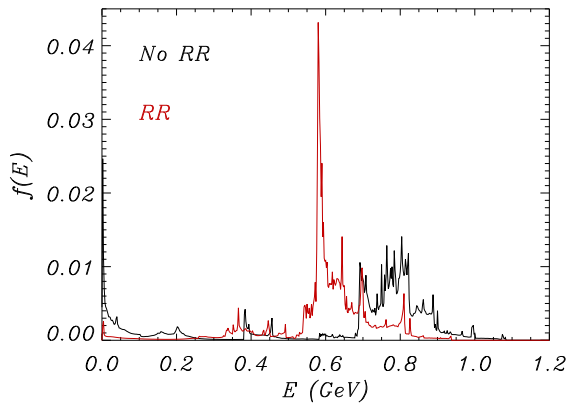}
\caption{\label{LPspectrum1} Ion energy spectrum at 
$t = 12 \, (T)$
with (red) and without (black) RR for LP and $I = 2.33 \times 10^{23} \Wcm$.}
\end{figure}

\begin{figure}[t!h]
\includegraphics[width=0.48\textwidth]{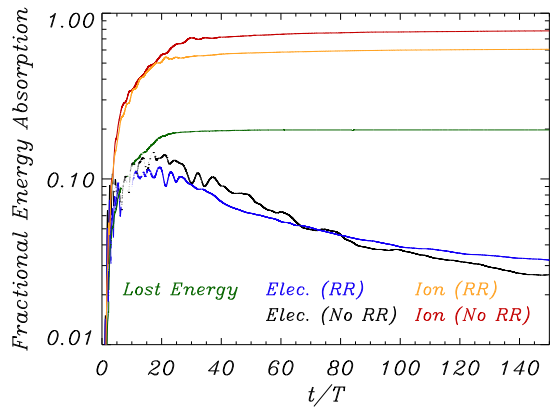}
\caption{\label{LPabs} Fractional energy absorption in function of time for LP 
and intensity $I = 2.33 \times 10^{23} \Wcm$. 
Electron kinetic energy with RR (blue)
and without RR (black), ion kinetic energy with RR (orange) and without RR (red)
and the fraction of energy lost in system (green).}
\end{figure}

For LP, the ion energy spectrum is significantly affected by RR effects.
The spectrum is fairly peaked with a \emph{smaller} energy spread and 
\emph{lower} peak energy than in the case without RR (Fig.~\ref{LPspectrum1}).
In general, as observed in many simulations the spectral peak produced 
by RPA broadens with increasing electron ``temperature'', since hot electrons
drive the expansion of the plasma leading to additional, non-monoenergetic ion
acceleration. The smaller energy spread observed when RR is included can be 
thus traced back to the radiative cooling of the most energetic electrons.
Moreover, a significant fraction of ions on the low energy tail 
of the spectrum is observed without RR, but disappears when RR is included.
The fractional difference in the ion energy with vs without RR is
of the order of the fraction of the laser pulse energy that is ``lost'' 
as incoherent emission (Fig.~\ref{LPabs}).
For $I = 2.33 \times 10^{23} \Wcm$, about 20\% of the total pulse energy is
lost as incoherent radiation (Fig.~\ref{LPabs}).

When RR is ``switched off'', part of the ``skin''layer of the foil
is left behind and a significant fraction of ions is present
on the low energy tail of the ion spectrum (Fig.~\ref{LPspectrum1}).
To explain this effect, we first recall that in the first stage of RPA
two ion populations may be produced, corresponding to a coherently moving 
``sail'' and to a trailing ``tail'' \cite{macchiNJP10}. Ions in the tail will 
eventually remain behind the sail if their charge is neutralized by returning
electrons; otherwise, they will be accelerated by their own space-charge field
and may move to the higher energy side. 
When the foil is still non-relativistic in the laboratory frame, 
the RR force has larger values when the electrons 
counter-propagate with respect to the laser pulse and therefore the electron 
backward motion is strongly impeded when the RR force is included. 
This effect prevents an efficient neutralization
of the ion charge in the tail by returning electrons, explaining why a higher
number of low-energy ions is observed without RR.

\begin{figure}[t!h]
\includegraphics[width=0.48\textwidth]{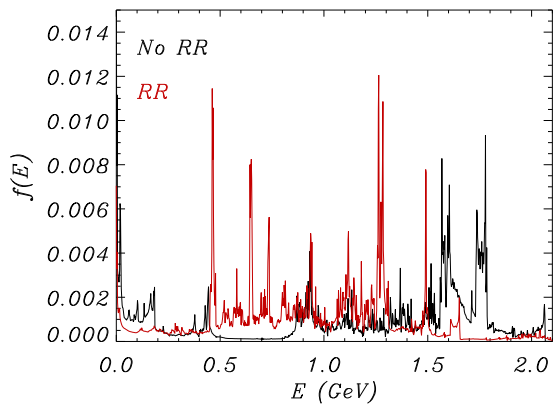}
\caption{\label{LPspectrum2} Ion energy spectrum at $t = 46 \, (T)$
with (red) and without (black) RR for LP and $I = 2.33 \times 10^{23} \Wcm$.}
\end{figure}

Equation (\ref{redLL}) suggests that the RR force is mainly a
nonlinear friction force. For $I = 2.33 \times 10^{23} \Wcm$,
about 20\% of the total pulse energy is ``dissipated''
by the RR force during the laser-foil interaction (Fig.~\ref{LPabs})
which lasts about 22 cycles (30 cycles without RR).
As stated previously, such ``dissipated'' energy accounts for
the incoherent radiation escaping from the plasma.
During the laser-foil interaction, such flux of incoherent radiation
shows itself in a missing pulse energy while ions
have almost the same total energy in both cases and
their spectrum is quasi-monochromatic (Fig.~\ref{LPspectrum1}).
However, after the acceleration phase by the radiation pressure of 
the laser pulse,
a 20\% of missing pulse energy implies about the same amount
of missing final ion energy (Fig.~\ref{LPabs}).
Moreover, a significant fraction of hot electrons are produced
by the $\bma{J} \times \bma{B}$ force. Such electrons can drive an expansion
of the foil, strongly increasing the ion energy spread
after the laser-foil interaction phase (Fig.~\ref{LPspectrum2}).

We remark that just changing the laser polarization from CP to LP,
the friction coefficient $d$ increases by up to \emph{four-orders}
of magnitude due the enhanced laser pulse penetration in the foil
by the $\bma{J} \times \bma{B}$-driven longitudinal oscillations.
Then, the electrons move in a strong electromagnetic field
becoming ultrarelativistic and the ``friction'' term of the
RR force becomes non negligible.
These results are a relevant test of the conjecture in 
Ref.\cite{esirkepovPRL04} that RR effects would be weak 
as the foil motion becomes relativistic. 
Our simulations suggest that this picture strictly holds only in the CP case,
where almost all of the foil moves at relativistic speed in the same direction 
of the laser pulse. In the LP case, a substantial fraction of electrons has 
both an ultrarelativistic motion in the transverse direction and a strong
oscillatory motion in the longitudinal direction, 
leading to significant RR effects.

The dependence of RR effects on the pulse polarization was also studied
in Refs.\cite{naumovaEPJD09,naumovaPRL09,schlegelPP09} for thick targets 
(``hole boring'' regime of RPA) and long pulse durations. It was also found that
RR effects are stronger for LP, although they are not negligible for CP
\cite{schlegelPP09}. These results cannot be compared straightforwardly to our
findings because of the quite different laser and plasma parameters, leading to
a different dynamics. For instance, in the thick target case  the laser-plasma 
surface oscillates also for CP (``piston oscillations'' \cite{schlegelPP09}) 
and a return current of electrons counterpropagating with respect to the laser 
pulse is generated; this effect is likely to enhance radiative losses.

\section{Conclusions}

We summarize our work as follows.
Radiation Reaction effects on  Radiation Pressure Acceleration of plasma
slabs by ultraintense laser pulses were studied by one-dimensional PIC 
simulations. The RR force was included via the Landau-Lifshitz approach.
The numerical implementation allows the addition of RR effects to any 
PIC code based on the standard Boris pusher algorithm for the acceleration 
of the particles, at a small computational cost.

We compared results for Circular and Linear Polarization of the laser pulse.
For CP, we found that RR effects become relevant only for plasma 
targets thin enough to let the laser pulse
break through the foil. In this case the inclusion of RR effects leads to an 
increase of the ion energy. Such increase is however not very significant 
with respect to a case with the same laser parameters but a thicker target, for
which the breakthrough of the laser pulse does not occur and RR effects are 
negligible.

For Linear Polarization, we found that RR effects are significant, leading to 
some tens of percent of energy loss by incoherent emission and to a reduction
of the peak ion energy by a similar percentage. Although RR effects produce a
somewhat more peaked energy spectrum during the acceleration stage, the final
spectrum is anyway dominated by a post-acceleration evolution, presumably 
driven by high energy electrons.

\section*{References}

\bibliographystyle{unsrt}
\bibliography{paper}

\end{document}